\documentclass[preprint2]{aastex}

\usepackage{natbib}
\usepackage{wasysym}
\usepackage{txfonts}
\def\AV{\mbox{A$_{\rm V}$}}

\def\WCO{\mbox{${\rm W}_{\rm CO}$}}         
\def\NCO{\mbox{${\rm N}_{\rm CO}$}}         
\def\XCO{\mbox{${\rm X}_{\rm CO}$}}         
\def\zXCO{\mbox{${\rm X}_{\rm CO}^0$}}         
\def\W13{\mbox{${\rm W}_{\rm 13}$}}

\def\nH2{\mbox{${\rm n}(\HH$)}}
\def\enH2{\mbox{$n_{(\HH$)}}}

\def\pccc{~{\rm cm}^{-3}} 
\def\ccc{~{\rm cm}^3} 
\def\pcc {~{\rm cm}^{-2}}

\def\Tsub#1 {\mbox{$T_#1$}}
\def\TK  {\Tsub K }

\def\fH2{\mbox{f$_\HH$}}
\def\mfH2{\mbox{$<{\rm f}_\HH>$}}

 \def\arcmin{\mbox{$^{\prime}$}}

\def\p{\mbox{$^+$}}

\def\h13cop{\mbox{{H$^{13}$CO\p}}}

\def\c3h2{\mbox{C$_3$H$_2$}}
 
 \def\R0{R$_0$}
\def\G0{\mbox{G$_0$}}

\def\ddeg{{}^\circ\kern-.1em}

\def\kms{\mbox{km\,s$^{-1}$}}
\def\ps{\mbox{~s$^{-1}$}}

\def\E#1 {$10^{#1}$}
\def\E#1 {E{#1}}
\def\P#1,{$\nH2\TK~=~#1\times~10^4\pccc$~K}
\def\ec#1,#2,#3,{#1\,(#2)\E{#3}}

\def\zoph{$\zeta$ Oph}

\def\H3{\mbox{H$_3$}}

\def\zetaH{\mbox{$\zeta_H$}}
\def\RH2{\mbox{R$_{\rm G}$}}
\def\GH2{\mbox{$\Gamma_{\HH}$}}
\def\g13{\mbox{g$_{13}$}} 
\def\thW{\mbox{W$_{13}$}}

\def\kHeH2{\mbox{$k_{ He-\HH}$}}

\def\tim#1,#2{\mbox{{$#1\times10^{#2}$}}}

\newcommand{\emm}[1]{\ensuremath{#1}}   % ensures math mode.
\newcommand{\emr}[1]{\emm{\mathrm{#1}}} % uses math roman fonts.
\newcommand{\hcop}{\emr{HCO^+}} 
 
\newcommand{\HH}{\emr{{\rm H}_2}}
\newcommand{\cotw}{\emr{^{12}CO}}
\renewcommand{\coth}{\emr{^{13}CO}}

\newcommand{\coei}{\emr{C^{18}O}}

\newcommand{\Kkms}{\emr{\,K\,km\,s^{-1}}}

\shorttitle{CO fractionation}
\shortauthors{H. S. Liszt }

\begin{document}

\date{generated \today}


\title{Formation and Fractionation of CO (carbon monoxide) in diffuse clouds observed at optical and radio wavelengths}


\author{H. S. Liszt}
\affil{National Radio Astronomy Observatory \\
            520 Edgemont Road,
           Charlottesville, VA,
           22903-2475}

\email{email to: hliszt@nrao.edu}



\begin{abstract}

We modelled \HH\ and CO formation incorporating the fractionation and selective 
photodissociation affecting CO when \AV\ $\la2$mag.  UV absorption measurements 
typically have N(\cotw)/N(\coth) $\approx 65$ that are reproduced with the standard UV 
radiation and little density dependence at n(H) $\approx32-1024\pccc$: Densities n(H)
$\la256\pccc$ avoid overproducing CO.  Sightlines observed in mm-wave absorption and a 
few in UV show enhanced \coth\ by factors of 2-4 and are explained by higher 
n(H) $\ga256\pccc$ and/or weaker radiation. The most difficult observations to understand 
are UV absorptions having N(\cotw)/N(\coth) $>$100 and N(CO)$\ga10^{15}\pcc$.
Plots of \WCO\ vs. N(CO) show that \WCO\ remains linearly proportional to N(CO) even at 
high opacity owing to sub-thermal excitation. \cotw\ and \coth\ have nearly the same 
curve of growth so their ratios of column density/integrated intensity are comparable 
even when different from the isotopic abundance ratio. For n(H)$\ga128\pccc$, plots of 
\WCO\ vs N(CH) are insensitive to n(H), and 
\WCO/N(CO)$\approx1\Kkms/(10^{15}~{\rm CO}\pcc)$: This compensates for small 
CO/\HH\ to make \WCO\ more readily detectable.  Rapid increases of N(CO) 
with n(H), N(H) and N(\HH) often render the CO bright, ie a small CO-\HH\ conversion factor.  
For n(H) $\la64\pccc$ CO enters the regime of truly weak excitation 
where \WCO $\propto$n(H)N(CO).
\WCO\ is a strong function of the average \HH\ fraction and models with \WCO=1\Kkms\ fall 
in the narrow range \mfH2\=0.65-0.8, or \mfH2\=0.4-0.5  at \WCO\=0.1\Kkms.   The insensitivity 
of easily-detected CO emission to gas with small \mfH2\ implies that even deep CO surveys using 
broad beams may not discover substantially more emission.

\end{abstract}


\keywords{astrochemistry . ISM: molecules . ISM: clouds. Galaxy}

%s1
\section{Introduction}

Carbon monoxide (CO) can form, be shielded and survive in detectable quantities 
inside even a very modest  H I-\HH\ transition in diffuse clouds at \AV\ $\le$ 1
mag where C\p\ is the dominant form of gas-phase carbon \citep{BurFra+10} 
and as little as a few percent of the total hydrogen column is in the 
form of \HH\ \citep{BurFra+07,SonWel+07,SheRog+08}.  In UV absorption, 
CO is detectable when N(CO) $\la 10^{12} \pcc $ at  
N(\HH) $\approx 10^{19}\pcc$ and  N(H) $\approx$ N(H I) + 2N(\HH) 
$\approx 2 \times 10^{20}\pcc$ \citep{CreFed04}\footnote{In this work
n(H) and N(H) are the total number and column densities of H-nuclei in all forms but
only H I and \HH\ are significant.}.  
Such small CO 
column densities are far below those at which mm-wave CO emission becomes 
detectable at typical sky survey sensitivities\WCO\ = 1 K-kms\ 
\citep{DamHar+01}, which is N(CO) $\approx 10^{15}\pcc$ \citep{LucLis98,Lis07CO}.

So it has long been possible to infer that mm-wave CO emission could not 
trace all of the \HH-bearing gas in the diffuse molecular interstellar medium
(ISM) where cool neutral atomic and molecular hydrogen coexist in appreciable
quantities.  How much of the diffuse interstellar molecular hydrogen exists 
in such regions is a question in its own right. It also bears on the question 
of ``dark'' gas generally, where the 
presence of more gas is indicated by gamma-rays or dust emission/extinction 
than would ordinarily be inferred from  21cm H I and mm-wave CO emission 
\citep{GreCas+05,Pla2011,Gre15Chameleon}.   
The gas shortfall has been variously attributed to optically thick H I 
that is underrepresented in H I emission
\citep[][but see \cite{StaMur+14}]{FukTor+15} or to \HH\ that is
missed in CO emission \citep{WolHol+10}.

Here we discuss the question of just what CO {\it emission} traces when the sightline
is in the diffuse/translucent domain (roughly, \AV $<$ 2 mag), where C\p\ is the 
dominant carrier of gas-phase carbon, the gas is somewhat warmer (30 - 80 K) 
but at typical thermal pressure p/k $\ga 2-3\times 10^3 \pccc$ K \citep{JenTri11} 
so that CO is weakly excited by ambient \HH\ \citep{Lis07CO} and, in some
cases, heavily fractionated \citep[][]{LisLuc98} owing to chemical 
isotope exchange \citep{WatAni+76,SmiAda80} and selective photodissociation 
\citep{BalLan82,vDiBla88,WarBen+96,VisVan+09}. 
These effects act simultaneously to produce isotoplogic abundance ratios that 
sometimes show the dominance of one or the other mechanism but may also result in 
normal-seeming abundance ratios that conceal the complex nature of the 
underlying physical processes.

%1
\begin{figure*}
\includegraphics[height=8.1cm]{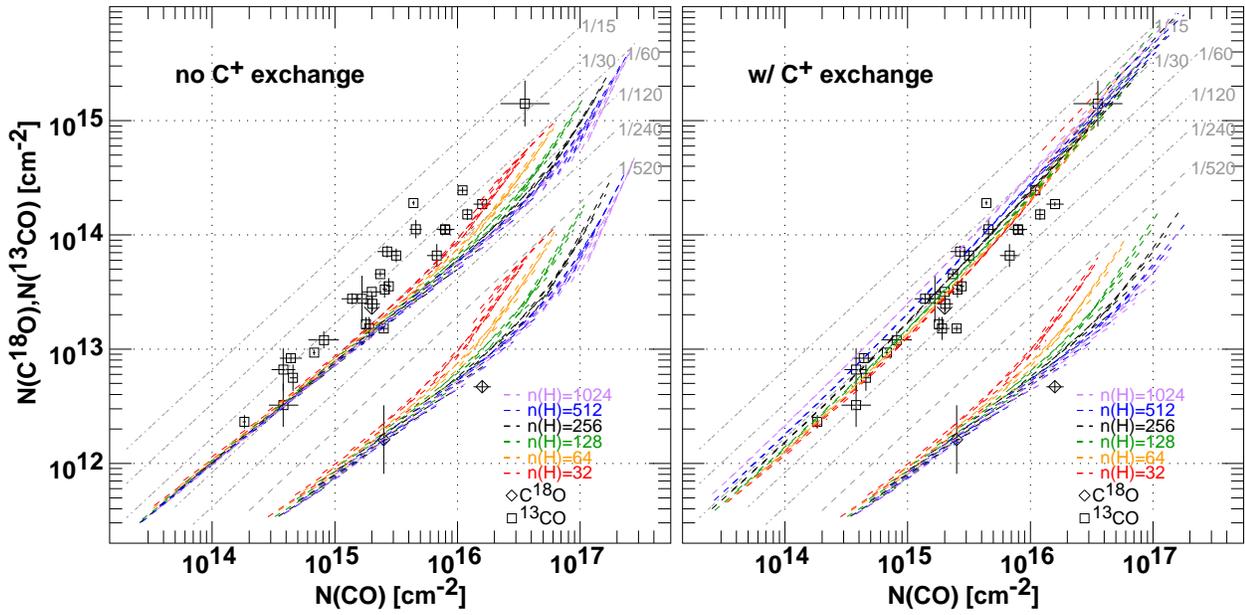}
  
\caption{Carbon monoxide column densities modelled with (right) and without 
  C\p\ exchange. In both panels
  N(\coth)\ and N(\coei) are plotted against N(CO) with UV-absorption
 observations shown as hollow rectangles and diamonds, respectively.  In
  each panel model results are shown for densities n(H) = 32 .. 1024$\pccc$
 varying in steps of two.  The calculations were taken from spherical clump models 
 in which the central column density N(H) was varied in steps of $2^{1/4}$ and
 for each model two values, corresponding to sightlines at the center and
 the geometric mean impact parameter, are connected by a dashed line . 
 In each panel, ratios with respect to the main isotope  are shown as shaded 
 dashed lines.}
\end{figure*}

%Interpretation of CO emission is somewhat peculiar  in this regime because the 
%integrated J=1-0 CO brightness \WCO\ is generally a measure of the CO column 
%density even when the J=1-0 transition is quite opaque, and the ratio 
%\WCO/N(CO) is relatively large, in both cases owing to the weak (strongly
%sub-thermal) excitation \citep{GolKwa74,Lis07CO}.  Moreover, the brightnesses
%of the isotopomers cannot be compared in conjunction with the inherent isotopic
%abundance ratios to derive the line optical depths, and in any case the CO abundance
%relative to \HH\ is small and expected to be  highly variable with small changes
%in the ambient density and UV-illimunation.

Interpretation of CO emission is quite particular in the diffuse/translucent
regime because the isotopologues can not be presumed to be present in the same 
proportions as the inherent atomic isotope ratios, so the line brightnesses 
of the isotopologues cannot be compared
with the inherent elemental abundance ratios to directly derive the line 
optical depths and column densities.  Even so, the CO column densities would be
of little use in inferring N(\HH) when most of the carbon is in the 
form of neutral atomic carbon or C\p, and the CO abundance relative to \HH\ is
small (ie $<< 10^{-4}$) and highly variable with respect to small changes
in the ambient conditions \citep{SzuGlo+16}.  At the same time, as we discuss
below,  sub-thermal excitation puts the J=1-0  rotational line in a radiative 
transfer regime where the isotopologic line brightnesses are proportional to 
their  respective column densities even for very optically
thick lines \citep{GolKwa74}, so the ratios of brightnesses reflect 
the ratio of abundance even when the observed values (say 15:1 for \cotw/\coth) 
would under other circumstances just be taken as evidence for heavily 
saturated \cotw\ emission. 

The plan of this work is as follows.  In Section 2 we describe the computational
devices used to calculate \HH\ and CO abundances and CO line brightnesses.  In 
Section 3 we compare models of the CO formation, excitation and fractionation with 
UV- and radio absorption line measurements where the most complete knowledge of \HH\
and CO column densities
is available.  In large part this is done in order to understand 
what is implied by the differences in fractionation that apparently
occur between sightlines toward bright early-type stars used in UV-absorption 
and those used in the mm-wave regime toward distant blazars.  But it also
serves the necessary purpose of benchmarking the models against the lines 
of sight where the most complete information on CO and \HH\ is available,
which seems necessary before discussing   In Section 4 we
discuss CO emission more generally, to correspond to the more usual case that only
CO emission is observed in one or more isotopologues.  
Section 5 is a summary and overview.

%s2
\section{Model calculations}

Discussion of CO formation and fractionation in diffuse \HH-bearing gas 
proceeds in several steps: i) description of the underlying physical 
properties of the host gas; ii) the \HH-formation and \HH\ self-shielding 
mechanism in the otherwise-atomic medium; iii) the formation chemistry 
of the carbon monoxide isotopologues and their photodissociation, 
self- and mutual shielding and chemical carbon isotope exchange 
\citep{WarBen+96,VisVan+09}; iv) calculation of the brightness
of the CO lines and the effect of CO radiation on the overall 
energy balance and temperature distribution: This is generally 
negligible here. 

\subsection{Heating, cooling and geometry}

As in our previous work, we adopt the heating-cooling model 
of \cite{WolMcK+03} at the Solar circle, as incorporated into a spherical cloud 
with uniform density.  The cloud is embedded in the ambient, isotropic, galactic 
radiation field of photons and cosmic-rays, with the default rate of the latter 
taken as $\zetaH = 2\times10^{-16}\ps$ per H-nucleus as seems appropriate for the 
diffuse molecular ISM \citep{McCHin+02,Lis03,HolKau+12,IndNeu+12,IndNeu+15}.  Most 
of the heating is from the photoelectron effect on small grains and
the cooling is through excitation of the fine structure lines of C\p\ and O I.
Rev2: The optical/UV radiation field is that of \cite{Dra78}, scaled
overall by a factor \G0\ whose default value is unity.
The equations of chemical and thermal balance are solved iteratively over a computational 
model with 128 or more equi-spaced radial shells, computing the radiation field in each 
shell averaged over the surrounding 4$\pi$ solid angle.  This formulation was recently 
used to study the formation and self-shielding of HD and \HH\ under conditions of
varying metallicity \citep{Lis15HD}.

\subsection{\HH\ formation and self-shielding}

Following the prescription of \cite[][see also \cite{SteLeP+14}]{Spi78} 
the rate constant for \HH-formation on grain surfaces is taken as 
\RH2\ $= 3\times 10^{-18}\ccc\ps \sqrt{\TK}$ where \TK\ is the locally-computed
kinetic temperature.  However, the thermal balance and temperature-dependent rate 
constant are not of crucial importance to the \HH-fraction.  Very nearly the same 
results are obtained using a fixed rate constant \RH2\ $= 3.9\times 10^{-17}\ccc\ps$ 
that is the average of the values obtained toward three stars by \cite{GryBou+02} 
and is often cited in other work, see \cite{Lis15HD}.  The most important point 
overall is to reproduce the observed mean \HH-derived kinetic temperature
$\approx$ 80 K  \citep{SavDra+77,RacSno+02} at the typical thermal pressures 
p/k $\approx 3000 \pccc$ K \citep{JenTri11} that are appropriate for 
the \HH-bearing regions. 

The models employ the \HH\ photodissociation and self-shielding scheme of 
\cite{DraBer96} which explicitly treats dust attenuation of the radiation field at 
the wavelengths of the Lyman and Werner bands of \HH\ (90 - 110 nm).
The optical depth for dust absorption is $\tau_d$ = $1.9\times 10^{-21}$N(H)
\citep{Dra03a}  as in \cite{SteLeP+14}. The free-space photodissocation 
rate of \HH\ is $4.25\times 10^{-11}\G0 \ps$, also as in \cite{SteLeP+14}. 
The accuracy of the \cite{DraBer96} formulation  was recently verified in 
great detail by \cite{SteLeP+14} using an exact calculation in the context
of the Meudon PDR code. Continuous absorption by dust is especially important
at densities n(H) $\la 32\pccc$ when large hydrogen columns are required
before much \HH\ accumulates \citep{Lis15HD}.

\subsection{CO formation}

The CO formation chemistry adopted here is extremely simple and direct:  CO forms
from the thermal recombination of a fixed quantity of \hcop\ at the locally-calculated
kinetic temperature; more explicitly, there is a constant assumed relative abundance 
X(\hcop) = n(\hcop)/n(\HH) = $3 \times 10^{-9}$ as inferred from observation
\citep{LisPet+10}.   The CO isotopologues are assumed to form in proportion to 
their inherent elemental abundance, which we take as \cotw:\coth:\coei\ 
= 1:1/60:1/520 \citep{LucLis98}.  
This reflects the fact that the isotopologues of \hcop\ should be present
in proportion to the inherent elemental isotopic abundance ratios, given the
difficulty of fractionating \hcop\ in diffuse gas.  \hcop\ recombines with an
electron many thousands of times faster than it reacts with an ambient \coth\ 
molecule,  which is the pathway that enhances H$^{13}$CO\p\ in dense,
fully-molecular clouds \citep{RouLoi+15}.

It is generally agreed that \hcop\ recombines to form the CO that is observed in 
diffuse gas \citep{VisVan+09} but the origin of the \hcop\ requires a mechanism that 
proceeds much faster than the rate of ion-molecule reactions in the gas that 
is modelled here.  As noted below it is one of the insights of this work that 
the bulk of the UV absorption line data and all of the observations of CO
absorption at mm-wavelengths are explained when the formation of CO from 
\hcop\ recombination and the in-situ fractionation and carbon isotope exchange 
of CO procede at thermal rates. 

\subsection{CO photodissociation}

The CO photodissociation scheme adopted here is that of \cite{VisVan+09}, numerically 
based on tables of shielding factors obtained from the website referenced there. 
The free-space photodissocation 
rates of the CO isotopologues are $2.6\times 10^{-10}\G0\ps$ for \cotw\ and
\coth, and $2.4\times 10^{-10}\G0 \ps$ for \coei.   The 
online tables have a much finer granularity (0.2 dex) than those published in the 
manuscript itself (1 dex).  In this scheme both dust and \HH\ shield the 
isotopologues 
but CO provides vastly more shielding for the rarer isotopologues than
the rarer isotopologues provide for themselves.  Therefore the tables use N(CO) 
as the independent variable for the shielding of all isotopologues.  The secondary 
effect of self-shielding of the rarer $^{13}$C-bearing isotopologues by themselves 
is accomodated by a 2nd table dimension that is the relative abundance of the rarer 
isotopologue, with the value 1:1/35 or 1:1/65; for the $^{18}$O-bearing isotopologues 
there is only one possibility, 1:1/560.  As discussed in Appendix A the shielding
factors of \cite{VisVan+09} are not always interpreted in this way. 

Because the calculations are not performed with the actual local relative isotopologic
abundances,  we experimented by repeating the same calculations with tables 
corresponding to the two carbon isotopologic abundance ratios.  In the unphysical 
case that C\p\ isotope exchange is neglected, the shielding table with \cotw/\coth\ 
= 65 is clearly preferable on physical grounds (see Figure 1) but there were only 
very slight differences of a few percent using the table with the higher 
\coth\ abundance.  When carbon isotope exchange is included (as it should be), 
use of both tables gives even much more similar results owing to the strong 
inter-species coupling.  In any case the small differences between the 
isotopologic ratios in the tables and the in-situ formation rate ratios in 
our work is of little consequence to the results that are presented.

\subsection{Carbon isotope exchange and other chemistry}

The process of exothermic (34.8 K) carbon isotope exchange in reactions of 
$^{13}$C\p with CO was introduced by \cite{WatAni+76}.  We adopt the  more
recent measured, temperature-dependent rate constants of \cite{SmiAda80} 
as parametrized by \cite{Lis07CO}, see  also \cite{RouLoi+15}.  Use of the
carbon exchange reaction in various implementations is discussed in Appendix A.

The formation/fractionation chemistry employed here was discussed in an approximate
way in \cite{Lis07CO}.  The important effects are CO formation, photodissociation, 
and destruction by He\p, along with the carbon isotope exchange.  We solved a limited 
chemical network consisting of these effects coupled with the underlying calculation
of ionization equilbrium and charge balance of all species that is inherent in the 
basic heating-cooling calculation.  In this way the equilibrium between carbon atoms, 
ions and CO is explicitly maintained for all the isotopes and isotopologues.

The model sphere calculation is performed radially in 128 equi-spaced shells using 
a relaxation mechanism that repeatedly recourses inward from the edge to the center 
until the abundances have converged in each shell and are self- and mutually 
consistent. The calculation is required to converge in the number densities of 
\HH, HD and the CO isotopologues, as well as the kinetic temperature
and ionization equilibrium, etc,  in each shell.  Once the internal 
radial variation of all quantities is known, the observable column densities and CO 
line brightnesses can be calculated for any impact parameter about the central 
sightline.  In the figures, we have typically calculated series of models at fixed 
number density n(H) while varying the front-back total central column density 
N(H) in steps of $2^{1/4}$, displaying the results for each model as a line 
segment connecting the results for 0 impact parameter and an impact parameter equal 
to 2/3 of the radius, which is the geometric mean over the face of the model.
As noted in the Introduction the model values for
n(H) and N(H) include H-nuclei in all forms, including protons, 
\HH\p, etc. but the fractional abundances of species other than atomic
and molecular hydrogen are very small so that 
N(H) $\approx$ N(H I) + 2 N(\HH).

\subsection{The CO J=1-0 emission line brightness and CO cooling}

As in \cite{Lis07CO} the carbon monoxide J=1-0 emission line brightnesses are
 calculated from a microturbulent approximation with complete redistribution 
\citep{LisLeu76} and CO cooling is also included in the model  although 
it is negliglible in the presence of so much C\p.  The rotational excitation 
of CO is implemented as described in \cite{Lis07CO} including excitation by 
atomic helium and hydrogen, ortho and para-\HH, 
although excitation by atomic hydrogen is actually negligible 
given the small rate constants \citep{SheYan+07,WalSon+15} and the absence of
CO except when the molecular hydrogen fraction is large.
The J=0 and 1 levels of \HH\ are taken to be in thermal equilbrium at the 
local kinetic temperature \citep{SavDra+77,RacSno+02}. 
The CO and \coth\ integrated line profile brightnesses in units of K \kms\ 
are written as \WCO\ and \W13\ respectively.

\subsection{Specific observational results discussed here}

Like \cite{VisVan+09} we discuss the CO and \coth\ column densities 
determined in UV absorption toward bright stars by \cite{SheRog+07} and 
\cite{SonWel+07} and the \coei\ column densities of \cite{LamShe+94}
and \cite{SheLam+02}.  We discuss the CO and \coth\ column densities
determined in mm-wave absorption by \cite{LisLuc98} and the emission 
brightnesses presented there and by \cite{Lis97} toward \zoph.  We also 
discuss the CO emission observed toward common UV absorption targets 
\citep{Lis08} and the ``Mask 1'' subset of CO emission brightnesses in 
Taurus observed by \cite{GolHey+08}.

\subsection{Comparison of approaches to shielding and isotope exchange with other work}

Similarities and differences in the self-shielding and carbon isotope exchange 
schemes used here and other work \citep{SheRog+07,RolOss13,SzuGlo+14,RouLoi+15,
SzuGlo+16} are discussed in Appendix A.

\subsection{A reference value for the CO-H$_2$ conversion factor}

Throughout this work we refer to the CO-\HH\ conversion factor as
\XCO\ = N(\HH)/\WCO\ and take as a standard or reference the value
\zXCO\ $=  2\times 10^{20}~\HH \pcc$/(K-\kms). 

%s3
\section{Comparison of absorption line observations with models}

%2
\begin{figure}
 \includegraphics[height=6.7cm]{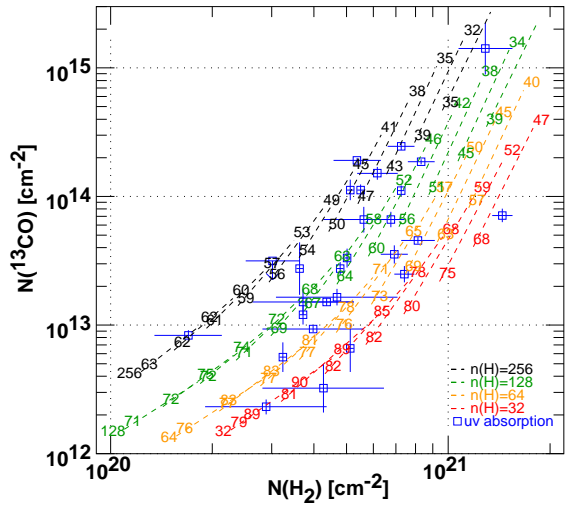}
  \caption{
N(\coth) vs N(\HH) for models shown in Figure 1 with the uv absorption
line data superposed as hollow blue rectangles.  As in Figure 1
each model is represented by a dashed line segment connecting the values at
the center and geometric median impact parameter, but in this figure the
N(\cotw)/N(\coth) ratios are shown at the endpoints of each line segment. 
The number density n(H) is indicated at the lefthand side of each series
of models. 
}
\end{figure}

%old 4 here 

%3
\begin{figure}
\includegraphics[height=6.3cm]{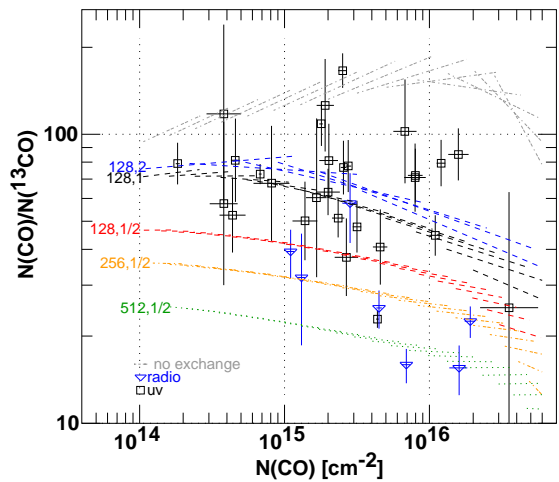}  
\caption{
The N(\cotw)/N(\coth) ratio for optical (hollow black rectangles) and radio
(blue downward triangles) absorption line observations, and models with
varying number density and scaled radiation field.  Each model 
calculation is labeled with its (n(H),G0) values. The dash-dot loci 
at the highest N(\cotw)/N(\coth) ratios correspond to calculations ignoring
carbon isotope exchange.}
\end{figure}

%4
\begin{figure*}
\includegraphics[height=7.1cm]{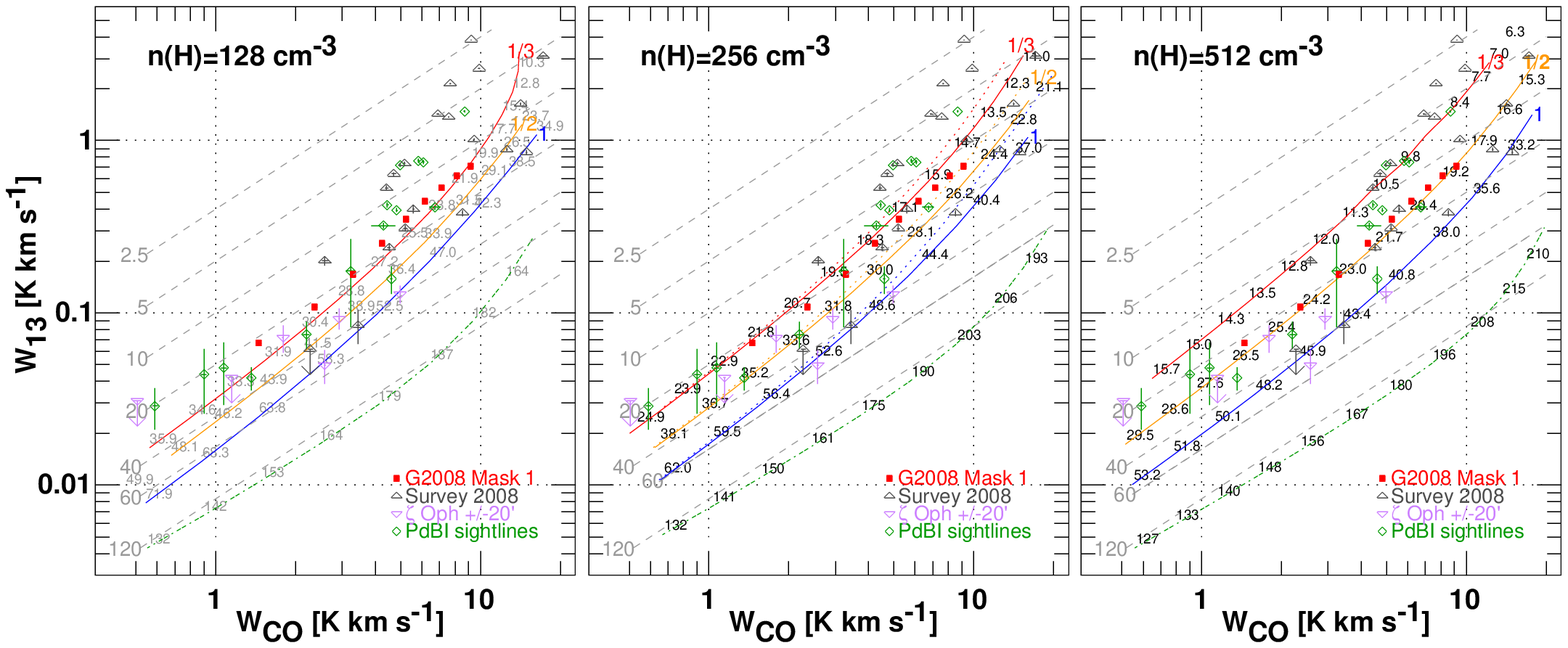}
\caption{
Emission profile integrals for \cotw\ and \coth\ and model results
for n(H)$=128\pccc, 256\pccc$ and $512\pccc$.  The datapoints include
sightlines toward early-type stars \citep{Lis08} (``Survey2008''), pointings
within 20\arcmin of \zoph\ \citep{Lis97}, PdBI sightlines toward compact extragalactic
continuum sources \citep{LisLuc98,LisPet+10} used for absorption line studies 
at the Plateau de Bure Interferometer (``PdBI sightlines'') and the ``Mask 1'' pixels
in the Taurus region discussed by \cite{GolHey+08}.  Denoting G0 as usual as the
ratio of the optical/UV radiation field to its standard value, models for 
G0=1,1/2 and 1/3
are shown in each panel as blue, orange and red lines, respectively,
and the  green dash-dot curve that is lowest in each panel shows model results 
neglecting isotope exchange. Gray dashed lines show fiducial values 
2.5, 5, 10 ... 120 of the \WCO/\W13\ intensity ratio and numbers along each
curve show the actual column density ratio N(\cotw)/N(\coth).
In the middle panel each model  is shown as a dotted curve with FWHM linewidth
dV = 0.9 \kms, dV = 1.3 \kms\ otherwise.}
\end{figure*}

Figure 1 shows the carbon monoxide column densities derived in UV absorption 
along with models with and without C\p\ exchange, at right and left, respectively.  
Note that all of the sightlines have N(CO) $ > 10^{14.1}\pcc$ and so reside
in the same regime of CO formation photochemistry initiated by processes
involving C\p\ and OH, according to \cite{SheRog+08}.
The  \cotw/\coth\ ratios are for the most part surprisingly 
well fit by the isotopologic abundance ratio N(\cotw)/N(\coth) =  C/$^{13}$C = 60
that is inherent in the model, while the observed N(CO)/N(\coei) ratios are 
three-five times higher than the intrinsic isotope $^{13}$C/$^{18}$O ratio 
520 used in the model.

The observed high  N(CO)/N(\coei) ratios are fit equally well with or without 
carbon isotope exchange because \coei\ is shielded mainly by \cotw\ whose
abundance is little changed.   The observed ratios 
N(\cotw)/N(\coth) $\approx$ 60 seem to ignore the expected complications of 
selective photodissociation, carbon isotope exchange and the like.  However, 
this is in no way an indication 
that carbon isotope exchange and selective  photodissociation are not occuring.  
The models shown at left ignoring carbon isotope exchange generally 
trace only the very lower envelope of the \coth\ data, especially at the 
lower densities, with  \coth\ column densities two-three times lower than 
observed.  The models without isotope exchange at left in Figure 1 
produce results very similar to those of the suprathermal chemistry employed 
by \cite{VisVan+09}, as discussed in Section 3.1. 

Introducing carbon isotope exchange at right in Figure 1 brings the model results 
into general agreement with the CO and \coth\ data, with little obvious dependence 
on the density, because the horizontal axis is N(CO) rather than, say, N(\HH) 
(see Section 3.1 and Figure 2). The upward curvature in the models is not apparent 
in the data unless the highest datapoint is considered. The improved fit at higher 
density for the \cotw/\coei\ ratio toward the star at higher column density occurs 
because N(H) and N(\HH) are both smaller at higher density for a given N(CO): 
the \coei\ is then less well shielded \HH\ and dust.

The diffuse \HH-bearing gas seen in UV absorption can be unambiguously distinguished
by its high inherent N(CO)/N(\coei) ratios that are three-five times larger than 
the inherent isotope ratio but distinguishing diffuse gas by comparing CO and 
\coth\  is far more problematical.  Deviations from 
the inherent isotopic abundance ratio are a clear sign of diffuse/translucent gas 
whether high or low  but only a small proportion of the \coth\ datapoints are 
strongly deviant.  

%where this belongs?
%In practical terms, modern radio receivers observe all three 
%isotopologues simultaneously so it is a simple matter to monitor all of them at 
%once and to integrate longer if the CO/\coth\ ratio turns out to be ambiguous.

\subsection{N(\coth)}

The CO formation rate is one of several factors competing to determine the 
\cotw/\coth\ ratio shown in Figure 1, so it is also important to reproduce the 
isotoplogic ratios at CO/\HH\ abundances that are also like those that
are actually observed.  If the CO abundance is not reproduced, conclusions
regarding the fractionation can be misleading.  Figure 2 shows  N(\coth) 
plotted against N(\HH) with the isotopologic abundance ratio shown at both
endpoints of each model's line segment (connecting the column densities calculated
toward the model's center and geometric-mean impact parameter).  

The N(\coth) and N(\HH) observed in UV absorption line data are reproduced 
at relatively modest densities $32\pccc \la$ n(H) $\la 256 \pccc$ as with 
the models in Figure 1.  The vertical scatter is much larger in Figure 2:
a factor 15-30 at given N(\HH) and with variation of N(\coth) by a factor
exceeding 100 for 
$3\times 10^{20}\pcc \la $ N(\coth) $\la 7\times 10^{20}\pcc$.  It can be 
explained  by variations in density and impact parameter as in Figure 2, 
but variations in the local UV illumination also must contribute.

%5
\begin{figure*}
\includegraphics[height=8.3cm]{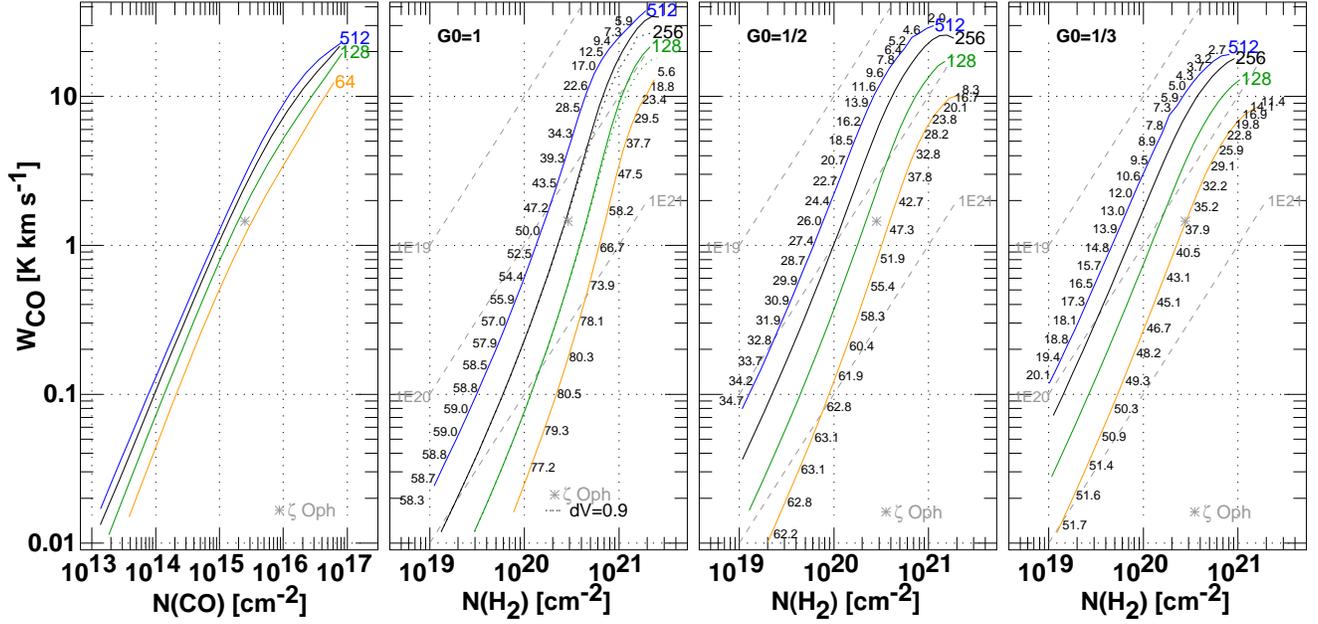}
\caption{ \cotw\ emission profile integrals \WCO\ plotted against \cotw\ and 
\HH\ column densities.
Numerical values along curves indicate the \WCO/\thW\ emission profile brightness
ratio.  The calculations used a FWHM linewidth dV = 1.3 \kms; calculations 
with dV = 0.9 \kms\ are shown as dotted lines for two densities in the panel
for G0 = 1.}
\end{figure*}

%6
\begin{figure}
\includegraphics[height=7.5cm]{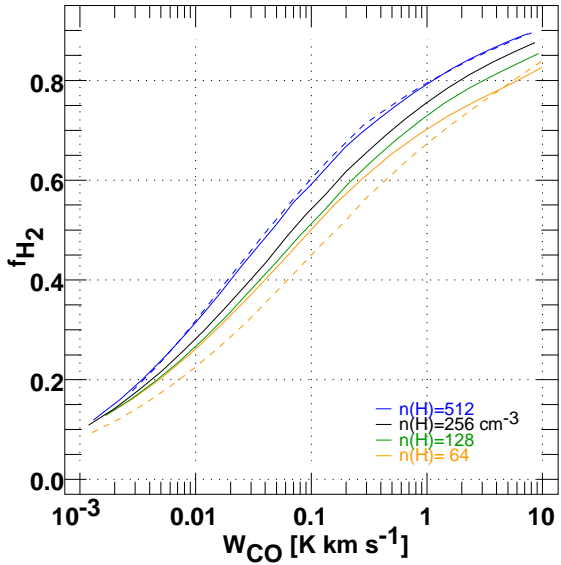}
\caption{
\HH-fraction plotted as a function of the CO profile integral toward the
center of models at four densities n(H) = 64 .. 512 $\pccc$.  Results 
are plotted as solid lines for all densities at G0 = 1 and as dashed lines 
for G0 = 1/2 at the highest and lowest densities.   The color 
scheme is the same as in Figure 1. }
\end{figure}

\subsection{Strongly deviant UV data and the radio-UV comparison}

In Figure 1 at right, there are three statistically significant datapoints 
with N(\cotw)/N(\coth) ratios well below the inherent carbon isotope ratio 
and five-six where N(\cotw)/N(\coth) is substantially larger. Deviations 
 also exist for the mm-wave data \citep{LisLuc98} but only in one sense and 
generally opposite to the UV data, with small N(\cotw)/N(\coth) and favoring chemical 
fractionation over selective photodissociation.  This is shown in Figure 3 
where the radio and UV data are compared: Other versions of this figure 
exist in \cite{Lis07CO}, \cite{SheRog+07}, \cite{VisVan+09}, and 
\cite{SzuGlo+14}, also using the results of \cite{LisLuc98}. With some overlap, 
the N(\cotw)/N(\coth) ratio is systematically smaller on sightlines 
observed at mm-wavelengths that do not specifically target early-type stars
and, albeit with a smaller sample, none of the radio datapoints show
N(\cotw)/N(\coth) exceeding the inherent elemental isotopic abundance ratio
of 60.

The radio or UV data with N(\cotw)/N(\coth) $< 60$ are easily accomodated with
modest (factor two) increases in density and/or smaller radiation fields.
However, increasing the radiation field or decreasing the density causes 
rather small changes in the N(CO)-N(\coth) curves and the most problematic 
data are those observed in the UV for which N(\cotw)/N(\coth) $> 80$.
Shown in Figure 3 as an upper envelope is the model result for 
n(H) $=128 \pccc$ neglecting carbon isotope exchange that is also shown at left 
in Figure 1.  The largest N(\cotw)/N(\coth) ratio, toward the archetypal line
 of sight to \zoph, is at the very margin of the values calculated ignoring carbon 
isotope exchange.  

Figures 1-3 allow a direct comparison 
with the chemical calculations of \cite{VisVan+09} from which our shielding factors 
were drawn.  Because \cite{VisVan+09} drove the isotope exchange reaction at 
suprathermal rates, with an effective temperature 4000 K, the 35 K zero-point 
energy difference between CO and \coth\ was ineffective, and their results
generally resemble those shown at left in Figure 1 here.
For instance, the earlier chemistry gave N(\cotw)/N(\coth) $\approx 140$ at 
N(\HH) $\approx 4-5 \times 10^{20}\pcc$ while the curves shown in our Figure 2
have N(\cotw)/N(\coth) $\approx 50-85$ over that same range.  The N(\coth)/N(\HH)
ratios in the earlier work are smaller than those shown in our Figure 2.

\subsection{Summing up; thermal and suprathermal chemistry}

Although our models including carbon isotope exchange cannot account for the 
handful of sightlines observed in UV absorption and having  N(\cotw)/N(\coth) 
$\ga 80$ at N(CO) $> 10^{15}\pcc$ (see Figure 3), effective carbon isotope 
exchange at thermal rates is needed to explain the bulk of the UV observations 
and all of the data at  mm-wavelengths.  The recombination of \hcop\ to form 
CO and the carbon isotope exchange affecting the N(\cotw)/N(\coth) ratio can 
be understood as occurring at thermal rates after the formation of \hcop, 
however that occurs.  As noted in Section 3 in reference to Figure 1, all 
of the sightlines considered here
in UV absorption have N(CO) $> 10^{14.1} \pcc$ and should be subject to the
same formation photochemistry according to the discussion of \cite{SheRog+08}.

As they discussed, high ratios N(\cotw)/N(\coth) occur as a matter of course in 
the suprathermal chemistry of \cite{VisVan+09}.  If the N(\cotw)/N(\coth) ratio 
is high because the carbon isotope exchange reaction were {\it ad hoc} being 
driven at suprathermal
rates with high effective temperatures 4000 K along some lines of sight, the 
recombination of \hcop\ to form CO would also presumably occur more slowly
owing to the inverse temperature dependence of recombination rates generally. 
For a given N(CO) this would in turn imply that X(\hcop) is higher in about the same 
proportion that its recombination rate to form CO is diminished.  

The \hcop\ column density cannot be measured directly along the sightlines 
studied in UV absorption, but the combination of higher N(\hcop) and 
somewhat stronger \hcop-electron excitation at higher temperature would 
presumably produce brighter \hcop\ rotational emission where the N(\cotw)/N(\coth) 
ratio was exceptionally large.  CO, \coth\ and \hcop\ emission have all been 
observed toward and around \zoph\ where the N(\cotw)/N(\coth) ratio is highest in 
UV absorption and the gas seen in emission is in 
the foreground of the star \citep{WilMau+92,KopGer+96,Lis97}.  Toward the star, 
\WCO\ $\approx 1.5$ K-\kms\ and \coth\ emission is at best marginally detectable 
with \WCO/\W13\ $\ga 60$ \citep{WilMau+92}.  \hcop\ emission is present 
at a level 2.2\% that of \WCO, typical of sightlines observed in \hcop\
absorption and emission in diffuse molecular gas at mm-wavelengths 
\citep{LucLis96}: It is not exceptionally strong.

CO, \coth\ and \hcop\ all brighten considerably within 10\arcmin-30\arcmin\ to 
the north and south of \zoph\ with the CO/\hcop\ brightness ratio staying about 
constant at 50-100 while \WCO/\W13\ decreases to 20-30.  
\hcop\ emission tracks \WCO\ fairly closely and varies with \WCO\ in the 
same sense as \W13.  This seems opposite to expectations if the suprathermal 
chemistry causes high effective kinetic termperatures in interactions of 
\hcop\ with electrons while retarding the carbon exchange reaction of 
C\p\ with CO.

%This is not observed along the sightlines
%observed at radio wavelengths where N(\hcop) is measured \citep{LisLuc96,LucLis96}
%but that is merely consistent with the lower N(\cotw)/N(\coth) ratios 
%that are determined at mm-wavelengths.

%s4
\section{CO emission in the face of fractionation }

In the vast majority of cases, observations of carbon monoxide
are those made in mm-wave emission without recourse to either absorption 
spectra or independent knowledge of N(\HH), etc: The integrated brightness 
of CO is often the independent variable or a constituent of it (eg when
N(H) is approximated as N(H) = N(H I)+ 2 \XCO \WCO).   Measurements 
of \coth\ emission in 
identifiably-diffuse gas are scarce but, in an attempt to show 
the equivalent of Figure 1 in emission,  Figure 4 shows measurements of CO 
and \coth\ emission from our work at the Plateau de Bure \citep{LisLuc98},
from a survey of carbon monoxide emission toward commonly-used UV absorption
background targets \citep{Lis08}, from data within 20\arcmin\ of \zoph\ 
\citep{Lis97} and from the lowest-\AV\ ``Mask 1'' pixels of 
\cite{GolHey+08} in Taurus.

The panels show results toward the centers of models at n(H) = 128, 256 and 
512 $\pccc$ with the radiation field at full strength and diluted by factors 
of 2 and 3. 
Each model curve is shown solid with FWHM = dV = 1.3 \kms\ and dotted in
the middle panel with dV = 0.9 \kms.  Fiducial values 2.5, 5, 10, 20 ...
of the \WCO/\W13\ integrated intensity ratio are  plotted as dashed gray lines,  
and dash-dotted in green we show the calculated emission when carbon isotope 
exchange is ignored for G0 = 1.  Numerical values of the column density
ratio N(\cotw)/N(\coth) are shown along each model curve:  These can be compared
with the fiducial lines to gauge the extent to which the intensity and
column density ratios differ owing to radiative transfer effects.   
Varying the density has relatively little effect on the model brightnesses
in Figure 4 when the radiation field is at full strength, consistent 
with the lack of density sensitivity shown in Figure 1 at right.  
For weaker illumination \WCO/\W13\ decreases with n(H), and 
moreso as the illumination diminishes.

The emission observations accompanying the PdBI absorption data are  
explained by some combination of higher density and dimmer radiation,
as in Figure 3 where the UV and mm-wave absorption line data were compared. 
The Taurus data ``Mask 1'' pixels of \cite{GolHey+08} are 
consistent with the other datasets, with $1/15 \le \W13/\WCO \le 1/30$.
It is interesting that the Taurus observations, which isolate diffuse 
sightlines within a dark cloud complex, are consistent with observations 
along sightlines observed at high latitudes at the PdBI that are supposedly 
remote from dense or dark gas. \coth\ emission  is often quite 
bright toward the early-type stars even though the UV absorption 
line data generally show smaller N(\coth)/N(CO) ratios than in the radio domain 
(ie Figure 3).  Strong carbon monoxide emission toward early-type stars 
almost certainly arises from dense material with high \AV\ situated behind 
the star.  Had the stars in question been located behind such material 
they would have too heavily extincted to be suitable absorption line targets.

\WCO/\W13\ intensity ratios are (perhaps) surprisingly close to the column
density ratios N(\cotw)/N(\coth) even when both are small compared to the intrinsic 
carbon isotope ratio 60: For \WCO\ $\la$ 4 K, the differences between the 
column density and brightness ratios are 25\% or less.
Differences are larger for brighter lines and for lower density, but
small intensity ratios reflect comparably small column density ratios, even when
\WCO/\W13\ $\approx 10 - 20$.  For instance the Taurus results with
\WCO/\W13\ in the range 13 - 22 are reproduced in the right-most panel
by the models with n(H) $=512 \pccc$, G0 = 1/2 and column density ratios
 N(\cotw)/N(\coth) in the range 19 - 26, or in the center panel at n(H)$=256\pccc$
with G0 $\ga$ 1/3 and N(\cotw)/N(\coth) in the range 16 - 22.  Close tracking of 
the intensity and column density ratios requires that the J=1-0 brightness maintains
a proportionality to the column density well beyond the point at which 
the optical depth exceeds unity, as we now discuss.

%As noted by \cite{SzuGlo+16} it is generally not profitable to attempt to derive 
%N(\coth) or N(CO) from such measurements in an attempt to extrapolate to N(\HH)

\subsection{\WCO\ vs. \NCO: The curve of growth}

Figure 5 shows the CO brightness plotted against N(CO) at left for G0 = 1 and
n(H) = 64, 128, 256 and 512 $\pccc$, and against N(\HH) at the same n(H) for 
G0 = 1, 1/2, and 1/3 in the three rightmost panels. The variation of \WCO\ with 
N(CO) haa little sensitivity  to the radiation field.
Values of \WCO/\W13\ are superposed numerically on the model curves in the 
rightmost panels where fiducial values of the CO-\HH\ conversion factor 
N(\HH)/\WCO\ in units of $\HH\ \pcc$/(K-\kms) are indicated as gray, dashed 
lines.

In the left-hand panel of Figure 5 the plots of 
\WCO\ vs N(CO) show that \WCO\ (K-\kms) $\approx$ N(CO)/$10^{15}\pcc$ around 
\WCO\ = 1 K-\kms\ for n(H) $\ga 128 \pccc$  and
\WCO\ $\approx$  N(CO)/$2\times 10^{15}\pccc$ for n(H) $= 64 \pccc$. 
These are comparatively low densities to be discussing, even in the context of
CO emission at the sensitivity limit of typical survey observations.
For n(H) $\la 64\pccc$, CO is increasingly in the true limit of weak collisional 
excitation where n(\HH)N(CO) is constant \citep{LisPet16}. 

When the excitation is weak and the  rotational energy ladder is well
below the level of thermalization, the J=1-0 brightness remains proportional 
to the CO column density long after the J=1-0 line becomes optically thick 
at with $\tau \ga 1$ at N(CO) $\ga 10^{15}\pcc$: This is because the radiative
transfer is dominated by scattering.  Moreover, the ratio 
\WCO/N(CO) is comparatively large in diffuse gas \citep{Lis07CO,LisPet+10}, 
some 30- 50 times higher than in cold dark clouds or when the rotational
ladder is thermalized, because so much of the energy in the rotation ladder
emerges in the 1-0 lowest transition.  This behaviour was predicted by 
\cite{GolKwa74} 
long before it was inferred from observation by direct comparison of \WCO\
and N(CO).  As discused in Section 5
a higher \WCO/N(CO) ratio compensates for smaller CO/\HH\ ratios in diffuse 
gas, such that \WCO/N(\HH) tends to remain constant between regions of high 
and low X(CO). The \WCO/N(\HH) ratio in diffuse gas differs much less than the 
\WCO/N(CO) ratio because the fractional abundance of CO is so small.  

%text from old obsolete appendix
 
%A.3 -8
%\begin{figure}
%\includegraphics[height=8cm]{Frac-FC2.eps}
%  \caption{The curve of growth for CO emission (left-most panel in Figure
% 5) on an expanded horizontal scale.} 
%\end{figure}

%%%%

\subsection{N(\HH)/\WCO : the CO-H$_2$ conversion factor} 

The right-hand three panels of Figure 5 show \WCO\ vs N(\HH) for G0 = 1, 1/2 
and 1/3 with  model results for n(H) = 64 .. 512 $\pccc$.  Casting the 
discussion in terms of \WCO\ and N(\HH) lends itself to discussion of the 
\WCO-N(\HH) conversion factor which is schematically illustrated in the figure
by the gray dashed fiducial lines.

The models in Figure 5 have an approximately quadratic dependence 
\WCO\ $\propto$ N(\HH)$^{1.5-2}$ and N(\HH) $= 2.1\times 10^{20}\pcc$ 
around \WCO\ = 1 K-\kms\ for G0=1 and n(H) = 256 $\pccc$.  \WCO\ varies more rapidly 
than linearly with N(\HH) at fixed number density owing to the fast increase of 
N(CO) with N(\HH) and this rapid increase of \WCO\ with N(\HH) means that CO-\HH\ 
conversion factors \XCO\ $<$ \zXCO\ often apply when \WCO\ $\ga 1$ K \kms.  
This is ``bright' CO \cite{LisPet12}, and CO is increasingly bright in
this sense at higher density and for smaller G0.

The CO-\HH\ conversion factor \XCO\ = N(\HH)/\WCO scales directly with G0 
(\XCO/G0 $\approx$ constant), and inversely with density
(\XCO\ $\propto$ n(H)$^{-0.75}$) and with \WCO\ itself, 
\XCO $\propto 1/\WCO^{0.3-0.5}$ with a slightly  stronger dependence
on \WCO\ at higher G0. The variation of \WCO\ with  N(\HH) shown in Figure 5 can 
be approximated around \WCO\ = 1 K-\kms\ and inverted to give
$$ N(\HH) = A y^b \times \WCO^{1/{(c+d{\rm ln}(y))}} \eqno(1) $$
%{\bf Check sign of d-term in Table 1}
where y = n(H)$/256\pccc$, \WCO\ is measured in K-\kms\ 
and constants for the evaluation of Equation 1 are given in Table 1.
The functional dependence is very nearly as G0/n(H) that is the fundamental 
scaling parameter for PDR in general. 

%t1
\begin{table}
\caption{Constants for the evaluation of Equation 1}
{
\begin{tabular}{lcccc}
\hline
G0 & A &b &c &d \\
\hline
1 & 2.24 $\times 10^{20}$ \HH\ (K-\kms)$^{-1}$ & -0.74 & 1.92 & -0.35 \\
 1/2 & 1.0  $\times 10^{20}$ \HH\ (K-\kms)$^{-1}$ & -0.79 & 1.66 & -0.12 \\
 1/3 & 0.69  $\times 10^{20}$ \HH\ (K-\kms)$^{-1}$ & -0.76 & 1.51 & -0.08 \\
\hline
\end{tabular}}
\end{table}

%The implied \WCO-N(\HH) conversion factor originally derived for the Mask 1 
%pixels was 
%N(\HH)/\WCO\  $= 2-10 \times 10^{20}\pcc$ (K-\kms)$^{-1}$ decreasing with 
%\WCO. Ratios N(\HH)/\WCO\ $ > 2 \times 10^{20}\pcc$ (K-\kms)$^{-1}$ are 
%emblematic of ``dark CO'', whereby  the presence of \HH\ is underrepresented in CO 
%emission.  Our models fitting the Taurus data generally show opposite behaviour,
%with N(\HH)/\WCO\ $ \la 5\times 10^{19}\pcc$ (K-\kms)$^{-1}$.  

In discussing Figure 4 we noted that our models with G0 = 1/2-1/3 and n(H) = 256 
$\pccc$ or G0 $\approx$ 1/2 and n(H) = 512 $\pccc$ reproduce the \WCO\ and \W13\  
measurements from which N(CO) and N(\HH) were derived for the Mask 1 pixels 
in Taurus of \cite{GolHey+08}.  
%The derived Mask 1 results are shown in Figure 6 and it can be seen that 
Our model with G0 = 0.5 and n(H) = 512 $\pccc$ in Figure 5 has the 
same CO-\HH\ conversion factor but five times smaller N(CO) and N(\HH) than were 
derived by \cite{GolHey+08}.  This occurs because the gas in our models 
is substantially warmer and the excitation more strongly sub-thermal than 
was assumed by \cite{GolHey+08}, which brightens the CO emission on a 
per-molecule basis while reproducing the observed \WCO\ with a smaller
CO column density, as discussed by \cite{LisPet+10}.
 
\subsection{Practical limitations on the utility of CO emission as an \HH\ tracer}

Figure 6 gathers results for models of varying density and G0 and plots the 
line of sight averaged molecular hydrogen fraction \mfH2\ against \WCO.  At the 
typical survey limit \WCO\ = 1 K-\kms, the models cluster in the narrow range 
\mfH2\ = 0.65 - 0.8, large but still with a significant component of atomic
hydrogen, falling only to \mfH2\ = 0.45 - 0.6 at \WCO\ = 0.1 K-\kms.  This suggests
that the known CO emission arises in material where the hydrogen has mostly 
been converted to molecular form and that even much deeper CO searches 
will not reveal much more emission except perhaps at higher spatial resolution.

This is not to say how much \HH\ is missed by CO surveys at the current
$\approx$ 1 k-\kms\
sensitivity limit, because that depends on the distribution of cloud properties
\footnote{Note also that this discussion assumes that the emission is 
resolved spatially, and that more stringent limits are implied otherwise} .  
Most of the \HH\ could reside in regions with a high enough molecular fraction 
that CO is detectable, though we noted earlier that there could be a substantial 
population of H I clouds with high column density and low number density and 
molecular fraction \citep{Lis07}.  Global estimates of \mfH2\ in the diffuse 
ISM based on absorption line observations range from $\ga 22$\% measured 
in \HH\ and corrected for sampling bias \citep{SavDra+77,BohSav+78},
to 35-40\% assuming constancy of the observed mean chemical abundances 
X(\hcop) $= 3\times 10^{-9}$ \citep{LisPet+10} and X(CH) $=  3.5\times 10^{-8}$
\citep{LisLuc02}.  In any case, the associated CO is not inferred to be 
abnormally dim \cite{LisPet+10}.  

%s5
\section{Summary and discussion}

We began by describing in Section 2 a physical scheme for numerical calculation
of the abundances of the carbon monoxide isotopolgues \cotw, \coth\ and \coei\ in 
uniform density gas spheres of modest total extinction and densities n(H) = 32 - 
1024 $\pccc$.  The models self-consistently compute the equilibrium state
considering heating and cooling of the host gas along with the \HH\ formation, 
self- and dust 
shielding in the otherwise-atomic medium; CO formation via the thermal recombination 
of a fixed relative abundance X(\hcop) = n(\hcop)/n(\HH) $= 3\times 10^{-9}$; and
CO photodissociation, self- and mutual shielding and carbon isotope exchange.
The emergent line intensities of the CO J=1-0 transition are calculated
in the microturbulent approximation.

In Section 3, as a benchmark, we compared the model calculations with UV 
observations of CO in absorption.  Figure 1, displaying results for models 
with and without carbon isotope exchange, shows that the observed isotopologic 
ratios N(\cotw)/N(\coth)
are generally well-reproduced over the full range N(\cotw) $= 2\times 10^{14} -
4 \times 10^{16} \pcc$ and that the near-constancy of N(\cotw)/N(\coth) $\approx$ 
65 at N(\cotw) $\la 2\times 10^{16} \pcc$, near the intrinsic $^{12}$C/$^{13}$C 
ratio, in fact results from a complex interaction of shielding, selective 
photodissociation and carbon isotope exchange processes.  When carbon isotope 
exchange is considered, the model results inter-comparing the various CO 
isotopologues  have little 
dependence on number density although with some preference for the highest density 
to explain one of only two datapoints for the rarer isotopologue \coei.

Because the isotopologic abundance ratios depend on the proper balance between 
formation, destruction and exchange processes we took care to verify that the 
CO isotopologic abundance ratios were reproduced at the observed column densities 
(Section 3.1 and Figure 2).  The number density should be moderate,
n(H) $\la 256 \pcc$, to avoid over-producing CO along sightlines observed in 
UV absorption and to maintain consistency with the rotational excitation 
temperatures that are an indirect by-product of measurements of CO absorption
in the UV \citep{SmiSte+79,CruWat81,WanPen+82,CreFed04,Lis07CO,Gol13}.  At such 
densities CO acounts for 
5 - 20\% of the gas-phase carbon in the center of models with 
N(H) = $1 - 2 \times 10^{21} \pcc$ (\AV\ $\approx$ 1/2 - 1 mag) as discussed
in Appendix B and illustrated in Figure B1.

Some deviant datapoints were discussed in Section 3.2.  Smaller 
N(\cotw)/N(\coth) ratios can be explained by weakening the 
UV-illumination by factors of 2-3 and by increasing the density, 
but the converse is not true: High ratios N(\cotw)/N(\coth) $ > 80$ are 
not reproduced by increasing the strength of the illumination.  
In Figure 3 we compared
the results for CO absorption observed at optical and radio wavelengths,
noting that the radio data have systematically smaller N(\cotw)/N(\coth) ratios.
This indicates that that they arise in conditions involving some combination 
of higher density and weaker UV illumination than is characteristic of the
foreground material observed toward early-type stars.  

High N(\cotw)/N(\coth) ratios, in the range 100 - 160, are seen only along
sightlines used to study UV absorption.  As shown in Figure 3, the 
largest N(\cotw)/N(\coth) ratios can be reproduced by suppressing 
the carbon isotope exchange, as occurs in  suprathermal chemistries like 
that of \cite{VisVan+09}.  In Section 3.3 we argued that suppressing carbon 
isotope exchange via a suprathermal chemistry would imply slower \hcop\
recombination to CO, requiring a larger N(\hcop)/N(CO) ratio and brighter \hcop\ 
mm-wave emission relative to that of CO.  However, observations of CO, \coth, and
\hcop\ emission toward and around \zoph, where the N(\cotw)/N(\coth) is
highest in the UV, show a more or less constant \hcop/CO brightess ratio 
around the star, even as \WCO/\W13\ falls to 20-30.

In Section 4 we turned attention to CO emission on its own terms, to discuss
the most common case where only emission from one of more
isotopologues is available and independent information on, for instance,
N(\HH) is not available.  In Figure 4 we showed  observations
of \cotw\ and \coth\ emission in identifiably-diffuse directions and compared
them with models at n(H)  $=128 - 512 \pccc$ and G0 = 1/3 - 1.
The CO emission data toward the same stars used for UV absorption are often
dominated by background gas having relatively strong \coth\ emission, because
of selection biases against high foreground extinction.  Observations toward
outlying portions of the Taurus cloud are consistent with the other datasets
(an important point given the supposed absence of high-\AV\ material in the 
vicinity of the  sighlines studied in mm-wave absorption) and are 
well-explained with n(H) $ = 512 \pccc$
and twice weaker UV illumination or n(H) $ = 256 \pccc$ and three times weaker 
UV illumination.  The CO brightness and column density ratios are
comparable even when they are much different from the inherent atomic
isotope ratio:  Small \WCO/\W13\ ratios imply small N(\cotw)/N(\coth) ratios
and small CO/\HH\ abundance ratios rather than a saturation of the emission 
from the more abundant isotopologue.
 
In Section 4.1 we considered the CO curve of growth for CO emission,
the variation of \WCO\ witn N(CO) noting that the emission brightness per 
CO molecule around \WCO\ = 1 K-\kms\ is 
\WCO/N(CO) $\approx$ 1 K-\kms$/10^{15}\pcc$ for n(H) $\ge 128 \pccc$ or 
\WCO/N(CO) $\approx$ 0.5 K-\kms$/10^{15}$(n(H)/64$\pccc$)$\pcc$  
for n(H) $\le 64 \pccc$.  The \WCO/N(CO) ratio is relatively high in 
diffuse gas because nearly all
the energy put into the CO rotation ladder by collisions emerges
in the  J=1-0 transition.  Moreover, \WCO\ $\propto$ N(CO) even at
N(CO) $> 10^{15} \pcc$ where the J=1-0 line has $\tau \approx 1$, because the
CO gas is effectively a pure scattering environment.  The density
dependence in the \WCO/N(CO) ratio at n(H) $\la 64\pcc$ is a sign that 
 CO is entering the limit of truly weak collisional excitation formulated 
by \cite{LinGol+77} and \cite{LisPet16}.

In  Figure 5 at right we showed calculated results for \WCO\ vs. N(\HH), which 
allowed a discussion of the CO-\HH\ conversion factor N(\HH)/\WCO\ in Section 
4.2 where we parametrized the dependence of N(\HH) on \WCO\ in equation 1
and Table 1.  The rapid increase of N(CO) with density and N(\HH),
combined with the proportionality between N(CO) and \WCO, causes much
of the CO emission at levels \WCO\ $\ga$ 1 K-\kms\ to be 'bright' in
the sense of having a CO-\HH\ conversion factor N(\HH)/\WCO\ smaller
than the 'standard' value \zXCO\ $=  2\times 10^{20}~\HH \pcc$/(K-\kms).

%The high brightness per CO molecule in diffuse molecular gas
%compensates for a small CO abundance 
%which helps to keep the CO-\HH\ conversion factor more nearly the
%same in diffuse and dense molecular gas, \citep{LisPet+10} ie
%$$\XCO = \frac{N(\HH)}{\WCO} = \frac{N(\HH)}{N({\rm CO})} 
%\times \frac{N({\rm CO})}{\WCO}. \eqno(2) $$
%
%Inserting the mean relative CO abundance  $<$\XCO$> = 3\times 10^{-6}$ 
%observed in UV absorption together with the expected brightness per CO 
%molecule N(CO)/\WCO\ $= 10^{15} \pcc$/(K-\kms) one finds  a CO-\HH\ 
%conversion factor \XCO\ $\approx$ \zXCO\ according to Eq. 2.
%However, \WCO/N(CO) is 
%rather constant in diffuse molecular gas and N(CO)/N(\HH) varies strongly 
%with local conditions such as N(\HH), density n(H) and illumination.  
%Hence the CO-\HH\ conversion factor in diffuse gas must vary locally 
%and the gas tends to be CO-bright, ie to have \XCO\ $<$ \zXCO,
%when \WCO\ exceeds 1 K-\kms.

In Section 4.3 (see Figure 6) we discussed how molecular emission at the level 
\WCO\ = 1 K-\kms\ arises from gas in a narrow range of line-of-sight averaged
\HH-fraction \mfH2\ = 0.65 - 0.8, only falling to  \mfH2\ = 0.45 - 0.6 at 
0.1 K-\kms.  
Easily detectable CO emission does not occur in gas except where \HH\ is
the dominant form of hydrogen.  This probably implies  that CO searches at 
levels well below \WCO\ = 1 K-\kms\ with broad beams will not detect much 
more CO emission.  However, CO  emission is heavily clumped on arcminute and 
smaller scales and quite bright patches may well exist within broad beams 
lacking detectable emission at levels 1 K-\kms\ \citep{Hei06,LisPet12}. 

How much diffuse molecular gas has been missed by the existing CO surveys 
is a separate question but the fact that the average CO-\HH\ 
conversion factor in diffuse molecular gas is near-standard \citep{LisPet+10}
indicates that most of the \HH\ in the diffuse ISM exists in regions
with sufficiently large local \fH2\ that CO emission should be detectable.
In any case it appears that detection of \HH\ in regions with very small 
\mfH2\ $<<$ 0.5 is best left to absorption line 
measurements of species like OH\p\ which actually prefer such  conditions 
and perhaps to observation of species like CH, OH and \hcop, whose abundances 
with respect to \HH\ are not as sensitive to ambient conditions as that
of CO \citep{GerNeu+16}.

% In fact, the most surprising 
%thing about the CO that is seen at \EBV\ $\la 0.2$ is how bright it may be 
%\citep{LisPet+10}.

\subsection{Discussion}

We discussed the optical/UV/radio absorption line data in order to
establish the suitability of the underlying models before extrapolating to cases 
where CO emission alone will be used to infer N(\HH) and few if any constraints 
are available.  To this end we note especially that the bulk of that data is 
understandable in terms of a chemistry that occurs at thermal rates,
after \hcop\ forms: specifically, this work invoked 
the thermal recombination of the observed amount of \hcop\ with electrons
to form the observed amounts of carbon monoxide, and the carbon isotope exchange 
that is required to explain \cotw/\coth\ ratios that are below or only modestly 
above the inherent isotopic abundance ratio.  Some UV sightlines do not entirely 
conform to this picture because their CO column densities are 
reproduced, at least approximately, even while N(\cotw)/N(\coth) has high 
values that only seem attainable if carbon isotope 
exchange is suppressed.  Overall, however, and toward all the sightlines
observed in absorption at mm-wavelengths, far from early-type stars and 
presumably more typical,  carbon isotope exchange 
fractionation dominates over selective photodissociation.  The great majority 
of the \coth\ observed in diffuse molecular gas arises through the deposition of 
$^{13}$C into CO, rather than the direct formation of \coth\ via the 
recombination of H$^{13}$CO\p.   This situation is directly comparable to 
that of the HD in the diffuse ISM, which results from the deposition of 
deuterons into existing \HH\ rather than the formation of HD on grains directly.

As shown in Figure 5 at left, the quantity that is most directly and reliably 
inferred from observations of CO emission in diffuse molecular gas is N(CO) 
because the emission brightness per CO molecule around \WCO\ = 1 K-\kms, 
\WCO/N(CO) $\approx$ 1 K-\kms$/10^{15}\pcc$, is insensitive to the illumination
of the gas, and insensitive to the number density when n(H) $\ga 128 \pccc$, or 
\WCO/N(CO) $\approx$ 0.5 K-\kms$/10^{15}\pcc$ (n(H)/64$\pccc$) for
n(H) $\la 64 \pccc$.
 
In this regime N(\cotw)/N(\coth) $\approx$ \WCO/\W13\ even when small values 
of \WCO/\W13\ might seem to imply substantial saturation of the CO emission 
with unfractionated column density ratios. 

%One test for the 
%presence of diffuse molecular gas would be to observe \coei:
%at G0 = 1, \coei\ would appear in emission at levels some three 
%times below that expected from \WCO\ and the inherent O/$^{18}$O ratio 
%(see Figure 1) and perhaps 5 times below that expected from \W13\ and the 
%intrinsic $^{13}$C/$^{18}$O ratio.  This test would be less effective
%under conditions of weaker illumination, as the weakly-shielded \coei\
%isotopologue is less subject to photoionization.

Failure to recognize that N(\cotw)/N(\coth) $\approx$ \WCO/\W13,
and that this implies that CO  carries a small fraction of the gas-phase
carbon,  can lead to serious errors when deriving molecular gas properties.  
Just this point has recently been addressed by \cite{SzuGlo+16} who concluded 
that the {\it least} reliable way to infer N(\HH) is to calculate N(\coth) and 
thereafter to compute N(CO) and N(\HH) without knowing the isotopologue ratio 
N(\cotw)/N(\coth) and CO abundance.  They conclude that use of a standard 
CO-\HH\ conversion factor applied to \WCO\ is more reliable.  In 
fact something very like this point was made long ago at a time 
when the CO-\HH\ conversion factor had not yet entered the mainstream and 
N(\HH) was determined by calculating N(\coth) in LTE at a temperature inferred 
from CO, and multiplying N(\coth) by the ratio appropriate for fully-molecular gas, 
N(\HH)/N(\coth) $\approx 5 \times 10^5$ \citep{Dic78}.  As noted then,
it is possible to arrive at a value for N(\HH) with much less effort and no
more uncertainty simply as N(\HH) = \XCO\ \WCO\ using a standard value for
\XCO\ \citep{Lis84}.

\acknowledgments

  The National Radio Astronomy Observatory is   operated by Associated
  Universities, Inc. under a cooperative agreement with the National Science 
  Foundation.  The hospitality of the ITU-R and Hotel Bel Esperance in Geneva
  and the UKATC and the Apex City Hotel in Edinburgh are appreciated during the 
  writing of this manuscript.


%% in parentheses, after the keyword, but they will not be verified.

%{\it Facilities:} \facility{}



\appendix

\section{Comparison of approaches to CO shielding and C\p\ exchange}

Recent work incorporates carbon monoxide photodissociation and/or fractionation in
chemical studies that cover rather different physical regimes from that
considered here, although with some consideration of results for CO column
densities measured in UV absorption as shown here in Figure 1.  
\cite{RolOss13} discuss carbon fractionation
chemistry in PDR with n(H) = $10^3 - 10^7 \pccc$ while \cite{SzuGlo+14}
discuss carbon monoxide chemistry in turbulent clouds of typical mass 
10$^4$ M$_{\rm Sun}$ and mean density n(H) = 300 $\pccc$, equivalent in
some rough sense to 
a uniform sphere of radius R=7 pc and central column density 
2n(H)R $ = 1.27 \times 10^{22}\pcc$ or \AV\ = 7 mag, well beyond the range
of column density considered here.

Unlike the current work, \cite{RolOss13} and \cite{SzuGlo+14} find noticeable
and very similar enhancement of $^{13}$CO at all values of N(CO) below
the point where all carbon is in CO, for instance with 
$15 \la$ N(\cotw)/N(\coth) $\la 40$ for $10^{13}\pcc \la$ N(CO) $\la 10^{17}\pcc$. 
As noted by \cite{SzuGlo+14} in regard to their Figure 14, equivalent to
our Figure 3, their results are much better suited to comparison with our
mm-wave absorption results \citep{LisLuc98} and the UV absorption
results must reflect different conditions than those that were considered.
 In the present work the difference is attributed to higher density and
weaker UV illumination along sightlines removed from early-type stars.

In fact these differences arise from differing treatment of carbon monoxide 
self-shielding and fractionation, as we now discuss.

\subsection{Self and mutual shielding}

\cite{RolOss13} used the shielding factors tabulated by \cite{vDiBla88} that
were calculated and tabulated in a form similar to those of \cite{VisVan+09} 
so their results would be most nearly comparable to those presented here.  
However, \cite{RolOss13}  reinterpreted the basic spectroscopic
scheme behind the shielding calculations that have been performed by van Dishoeck 
and her collaborators, in which the shielding of all the isotopologues 
$^xC^yO$ is expressed as functions of N(\HH) and N(\cotw), not of N(\HH) and 
N($^xC^yO$), etc.  The ratios  N($^xC^yO$)/N(\cotw) are secondary 
parameters of limited significance in the formulation of \cite{VisVan+09}.
%even though they properly should be considered and calculated self-consistently.
Thus footnote 11 of \cite{RolOss13} states that
``\cotw\ could also shield its less abundant relatives, if their absorption
lines are sufficiently close together. However, only very few of the efficiently
dissociating lines overlap, too few for mutual shielding to be important 
\citep{WarBen+96}.''  However,  the footnote to Table 5 
of self-shieldings in 
\cite{VisVan+09} makes explicit that ``Self-shielding is mostly negligible 
for the heavier isotopologues, so all shielding functions are expressed as 
a function of the \cotw\ column density.''  This reflects a fundamental 
disagreement between the microphysics of line overlap between \cite{WarBen+96} 
and \cite{vDiBla88} or \cite{VisVan+09}. 

The first preprint version of \cite{SzuGlo+14} that we downloaded when initially
drafting this manuscript adopted the same self-shielding formulation of 
\cite{LeeHer+96} that we used in our own earlier work when only \cotw\  
was discussed, and they modelled the 
shielding of \coth\ by substituting N(\coth) for N(\cotw): That is, both
isotopologues use the same shielding factors, but the column density
employed to calculate the shielding of $^x$C$^y$O was N($^x$C$^y$O)
as in the scheme of \cite{RolOss13}.  The final published version uses
the shielding factors of \cite{VisVan+09} but presumably with the
same scheme of self-shielding of the heavier isotopologues, 
rather than by \cotw.  This presumably also is true of the
successor work by \cite{SzuGlo+16}.  

%To understand the effect of this approach  we repeated some of the models
%whose results were displayed in our Figure 1 using such a formulation and the
%self-shielding scheme of \cite{LeeHer+96}, with results as shown in Figure 7.
%When carbon exchange is ignored and for \coei\ the scheme adopted
%by \cite{SzuGlo+14} provides stronger shielding of the heavier isotopologues 
%and is a somewhat poorer match for the \coei\ observations.  When C\p\ 
%exchange is enabled, comparison of the green curve in Figure 8 with that 
%in Figure 1 at right shows that the scheme adopted by \cite{SzuGlo+14} 
%yields N(\cotw)/N(\coth) $<$ 60 at all N(CO) while this only occurs in our 
%Figure 1 when N(CO) $\ga  2\times10^{15}\pcc$.  The scheme adopted
%by \cite{SzuGlo+14} also produces smaller N(\cotw)/N(\coth) at higher
%N(CO) and this is entirely consistent with the results shown in 
%\cite{SzuGlo+14}.

\subsection{The carbon-exchange reaction rate}

Yet another difference with the present work is that \cite{SzuGlo+14} (and, 
presumably \cite{SzuGlo+16}) used a single value for the forward
rate constant (ie that for the exothermic replacement of C by $^{13}$C in CO), 
as measured at 300 K by \cite{WatAni+76}.  In doing so they followed a remark 
by \cite{SheRog+07} to the effect that the older result provided a better 
fit to the UV absorption line observations, a few of which have high
N(\cotw)/N(\coth) ratios. We, along with \cite{RolOss13} (also see \cite{RouLoi+15}) 
approximated the 
temperature-dependent rate coefficients for C\p\ exchange measured by 
\cite{SmiAda80}.  These are higher than the value 
$k_f = 2 \times 10^{-10} {\rm cm}^3$\ps\ of \cite{WatAni+76} by factors 
of 2, 3 and 4 at at kinetic temperatures 160 K, 80 K, and 30 K, respectively.

%\subsection{Influence of the CO formation chemistry}
%
% The CO formation chemistry plays a large role in determining the isotopologic 
%abundance ratio \citep{Lis07CO}.  When the formation rate is large, it is an 
%rapid immediate source of carbon monoxide at the inherent C/$^{13}$C 
%ratio.  When the formation rate is low, $^{13}$C is more likely to be formed 

%A.1
%\begin{figure}
%\includegraphics[height=7cm]{Fractionate-FigA1.eps}  
%\caption{
% Observed and model CO isotopologue column densities in the presence of different
% shielding models at n(H) $= 128\pccc$.  The black curve labelled 
% ``this work, no exchange'' is the curve for n(H) $= 128\pccc$ in the left-side
% panel of Figure 1 neglecting C\p\ exchange.  The red and green curves 
% labeled ``Szucs'' employ the self-shielding model adopted by  
% \cite{SzuGlo+14} using the shielding factors of \cite{LeeHer+96}, 
% without C\p\ exchange in red and with it in green. }
%\end{figure}
 
\section{Internal structure}

Figure B1 shows the internal structure of our cloud models with 
N(H) = $1\times 10^{21}\pcc$ and  $2\times 10^{21}\pcc$ at top and 
bottom, respectively.  The CO column densities toward the centers of these
models vary from $8.1 \times 10^{13} \pcc$ to  $1.3 \times 10^{16} \pcc$
at top (a factor 160 vs. 8 in density) and from  
$8.8 \times 10^{14} \pcc$ to  $8.3 \times 10^{16} \pcc$
at bottom.  For comparison with Figures 1 and 3, the central CO column
densities at n(H) $= 128 \pccc$ are  $1.1 \times 10^{15} \pcc$ 
at top and  $1.1 \times 10^{16} \pcc$ at bottom.

%%%%%%%%%%%%%%%%%%%%%%5 was here

The temperature distribution is nearly uniform because the models are 
relatively transparent (\AV\ = 0.25 and 0.5 at the center)  and the CO 
fraction -- the CO abundance relative to the free gas-phase carbon 
abundance -- is generally small.  The temperature at the center of the 
densest model at bottom increases toward the center as C\p\ becomes
a minority constituent and CO cooling does not make up the difference.  
The outer temperature varies from 55 K to 100 K
as the density decreases from $512 \pcc$ to $32 \pcc$.  The thermal pressure
p/k $\approx 3\times 10^3\pccc$K at n(H) $= 32\pccc$ was a basic constraint
on the underlying heating-cooling calculation.  The thermal pressure of
the highest density models is 4 times higher than this at the outer edge 
where the gas is mostly atomic, and about twice as large  at the center where the
gas is mostly molecular.

The models mix observable CO with substantial atomic gas fractions; at 
n(H) $= 128 \pccc $ the H I fraction everywhere exceeds 
about 20\% at top and 9\% at bottom in Figure 3.  The CO fraction is by far the 
most widely-varying cloud property, from  0.1\% to 20\% at the center 
at top to 0.7\% to 50\%  at bottom.

The n(CO)/n(\coth) ratio may either increase or decrease toward the center
depending on the density and may be both above and below 60 in the same
model at intermediate values of the number density.  The CO fraction is not 
large enough in any of the models to cause the isotopologic ratio to revert 
to 60 anywhere; the selective deposition of $^{13}$C into CO continues even
when the CO fraction is as high as 40\% at the center of the densest model.
The interactions between physical processes are complex because CO is also 
providing the bulk of the shielding for \coth.
 
%A.2 - 7
\begin{figure}
\includegraphics[height=14.9cm]{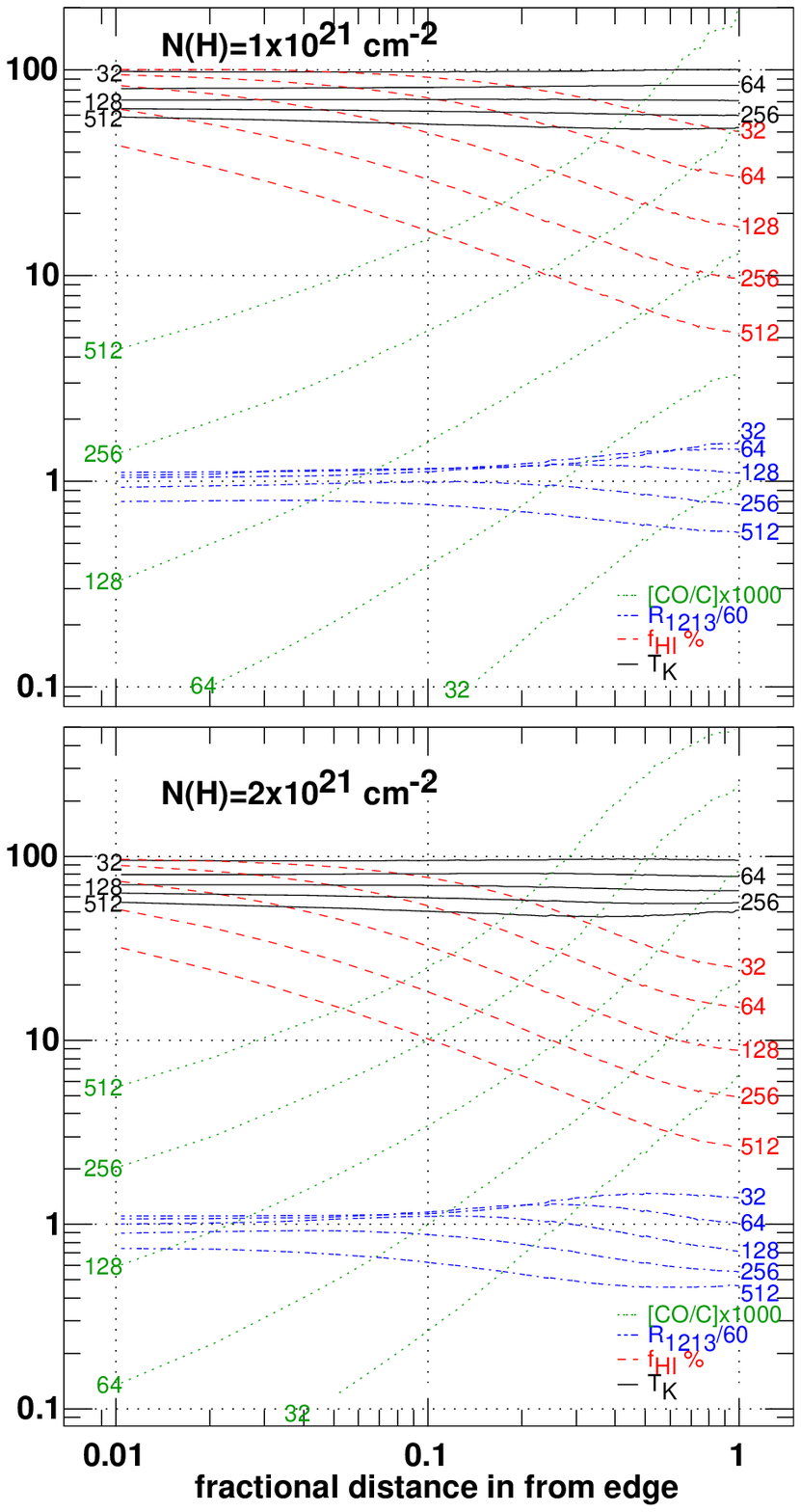}
  \caption{Radial variations in models with number densities 
n(H) = 32 .. 512 $\pccc$ and front-back column densities
N(H) $= 1\times10^{21}\pcc$ (upper) and N(H) $= 2\times10^{21}\pcc$ (lower).  
Shown are the kinetic temperture \TK\ in black and solid; the 
percentage fraction of hydrogen in atomic form f$_{\rm HI}$ in red, dashed; 
n(CO)/n(\coth) dash-dotted and blue; and $1000\times$ n(CO)/n(C\p) in green 
and dotted.} 
\end{figure}

%comment out next line to insert .bbl
%\bibliography{mnemonic,absorption}

\end{document}